\documentstyle[epsfig,aps]{revtex}

\def\gsim{ \,\, \vcenter{\hbox{$\buildrel{\displaystyle >}\over\sim$}}
 \,\,}
\def\lton{ \,\, \vcenter{\hbox{$\buildrel{\displaystyle <}\over\sim$}}
 \,\,}

\newcommand{\be}{\begin{eqnarray}}
\newcommand{\ee}{\end{eqnarray}}
\newcommand{\non}{\nonumber\\}
\newcommand{\bea}{\begin{eqnarray}}
\newcommand{\eea}{\end{eqnarray}}

\begin{document}

\thispagestyle{empty}
\title{\bf Proton breakup in high-energy $pA$ collisions from perturbative QCD}

\author
{
 A.~Dumitru$^a$, L.~Gerland$^b$, M.~Strikman$^c$\\
 {\small\it $^a$Physics Department, Brookhaven National Laboratory, Upton,
            NY 11973, USA}\\
 {\small\it $^b$School of Physics and Astronomy, Raymond and Beverly
Sackler Faculty of Exact Science, Tel Aviv University, Tel Aviv, Israel}\\
 {\small\it $^c$Department of Physics, Pennsylvania State University,
                University Park, PA 16802, USA}\\
}

\maketitle

\begin{abstract}
We argue that the distribution of hadrons near the longitudinal
light-cone in central high-energy
$pA$ collisions is computable in weak coupling QCD. This is because the
density of gluons per unit transverse area in the dense target at saturation
provides an intrinsic semi-hard momentum scale, $Q_s$. We predict that
the longitudinal distribution of (anti-)baryons and mesons
steepens with increasing energy and atomic number of the
target, and that the transverse momentum distribution broadens.
We show that the evolution of high moments of the longitudinal net baryon
distribution with $Q_s$ is determined by the anomalous dimension
$\gamma_{qq}$.
\end{abstract}
\vspace*{1cm}

The computation of leading hadron (baryon) production in inelastic hadronic
collisions from perturbative QCD (pQCD) usually suffers from the absence of
a large momentum transfer scale. For example, in $pp$ or $pA$
collisions (at not too high energy) many times the incident baryon stays
intact and emerges in the
final state with a large fraction $z$ of the projectile light-cone
momentum (the ``leading particle'' effect, see the discussion and further
references
in~\cite{ageev,dok}). Phenomenologically, one may picture this
as the interaction of only one of the valence quarks of the proton with the
target, which is then shifted to smaller $z$. The spectator diquark hadronizes
and carries away roughly $z\simeq
2/3$ of the incident momentum. Simple models like the ``inside-outside''
cascade assume dominance of such soft processes for leading baryon production
up to high energies and so predict that baryon number is largely
concentrated about the rapidity of the incident proton, and at small transverse
momentum. Nonperturbative models based on the ``gluon junction'' and
``diquark breaking'' mechanisms
have been proposed in the literature to describe the breakup of the
projectile, see e.g.~\cite{junction}.
In such models, also, baryons and anti-baryons are produced mainly at small
transverse momentum.

In this note, we argue however that the inclusive
distribution of leading hadrons in fact
can be computed in weak coupling for very high energy collisions when the
target approaches the ``black body'' limit, and that leading hadron
production is strongly suppressed. This is because for a
dense target all incident
proton constituents scatter and experience a large momentum transfer (which is
set by the ``saturation'' density scale of the target). Thus, the coherence of
the projectile is destroyed completely, and the scattered quarks and gluons
fragment independently~\cite{Berera:1996ku,dok,djm2,tl}.
As a consequence, the proton decays predominantly into a beam of
leading {\em mesons}, with the baryon number shifted to small light-cone
momentum fraction $z\lton0.1$. Also, we expect that the
$p_t$ distributions of leading baryons and mesons
closely reflect that of the scattered
quarks, which is rather flat up to transverse momenta on the order of
the square root of the density of gluons per unit area in the dense target.

The phenomenon we discuss here can be understood in two complementary
languages. The limiting case of scattering off a black body is more transparent
in the frame where the nucleus is at rest. The partons up to a resolution
scale $p_t(s)$ (which is somewhat higher for gluons than for 
quarks) interact with the target with the geometric cross section
of $2\pi R_A^2$. This implies the natural though pretty strong 
assumption that the parton - nucleus cross section
is not tamed at a substantially smaller value.
Hence in this limit the projectile wave function is resolved at a virtuality 
of $\sim p_t^2$ which is much higher than for soft processes. 
In this frame, the process of leading hadron production corresponds to
releasing the resolved partons from the projectile wave function.
Correspondingly, the partons fragment with large transverse momenta $\sim p_t$
and essentially independently since their coherence was completely 
lost in the propagation through the black body. In the case 
of $\gamma^* A $ scattering one is able to make model 
independent predictions for the leading hadron spectrum~\cite{bbl} which
differ drastically from the DGLAP limit, providing an unambigous signal for the
violation of the leading twist approximation.

A complementary way to discuss the limit of high densities is the 
infinite momentum frame treatment. Indeed,
the wave function of a hadron (or nucleus) boosted to large rapidity exhibits
a large number of gluons at small $x$.
The density of gluons is expected to saturate
when it becomes so large that gluon splitting is balanced by
recombination~\cite{mueller}. The
density of gluons per unit of transverse area and of rapidity at
saturation is denoted by $Q_s^2$, the so-called saturation momentum. This
provides an intrinsic momentum scale~\cite{sat}
which grows with atomic number (for
nuclei) and with rapidity, due to continued gluon radiation as phase space
grows. For sufficiently high energies and/or large nuclei, $Q_s$
can become much larger than $\Lambda_{\rm QCD}$ and so weak coupling methods
are applicable. Nevertheless, the well known leading-twist pQCD can not be used
precisely because of the fact that the density of gluons is large; rather,
scattering amplitudes have to be resummed to all orders in the density. When
probed at a scale below $Q_s$, scattering cross sections approach the
geometrical size of the hadron (the ``black body'' limit), while far above
$Q_s$ the hadron is in the dilute regime where cross sections are approximately
determined by the known leading-twist pQCD expressions.
For practical applications we actually need a model which interpolates
between the black body kinematics below the scale $Q_s$ and large $Q^2$
where pQCD is applicable. For these purposes we shall employ the
``Color Glass Condensate'' model of~\cite{sat}. 

We shall focus here on inclusive hadron production in the forward region of
central $pA$ collisions at high energies, such as at BNL-RHIC and CERN-LHC. The
``forward region'' is defined as the fragmentation region of the
projectile, which may also be a small nucleus rather than a proton. (In the
near future, deuteron-gold collisions will be studied experimentally at RHIC.)
Such central, small impact parameter collisions  can be selected using 
additional experimental triggers in the fragmentation region of the big
nucleus.

The target
nucleus, when seen from the projectile fragmentation region, is characterized
by a relatively large saturation momentum. Its precise value can not be
computed from first principles at present, but extrapolation of HERA
fits to nuclei including the gluon shadowing effects
yield values of $Q_s^2\sim10$~GeV$^2$ 
at $x\sim 10^{-4}$ reachable for BNL-RHIC energy and $\sim50$~GeV$^2$ 
at $x\sim 10^{-6}$ for CERN-LHC energy (for $A\sim200$ central
nuclear targets)~\cite{gbw}. For LHC
energies, it might even be interesting to consider head-on central $pp$
collisions: seen from the fragmentation region of one of the protons, the other
hadron might exhibit gluon densities as high as those for $A\sim200$ nuclear
targets at RHIC energy.

In the above-mentioned kinematic
domain, the dominant process is scattering of quarks from the incident dilute
projectile on the dense target. (In our numerical estimates we include
scattering of gluons, using the factorization theorem proven in~\cite{djm1}.
However, for leading hadron production, $z\gsim0.1$, this amounts to only a
small correction since
the gluon spectrum drops faster with $z$, and because gluon $\to$ hadron 
fragmentation is steeper.) At high energies (in the eikonal approximation),
the transverse
momentum distribution of quarks is essentially given by the 
correlation function of two Wilson lines $V$ running along the light cone 
at transverse separation $r_t$
(in the amplitude and its complex conjugate),
\be
\sigma^{qA} = \int \frac{d^2q_t dq^+}{(2\pi)^2} \delta(q^+ - p^+)
\left<\frac{1}{N_c}\,{\rm tr}\,
\left| \int d^2 z_t \, e^{i\vec{q}_t\cdot \vec{z}_t} \left[
V(z_t)-1\right] \right|^2\right>~.
\ee
Here, the convention is that the incident proton has positive rapidity, i.e.\
the large component of its light-cone momentum is $P^+$, and that of the
incoming quark is $p^+=x P^+$ ($q^+$ for the outgoing quark).
The two-point function has to be evaluated in the background field of the
target nucleus.
A relatively simple closed expression can be obtained~\cite{djm2} in the
``Color Glass Condensate'' model of the small-$x$ gluon distribution of the
dense target~\cite{sat}.
In that model, the small-$x$ gluons are described as a stochastic
classical~\cite{classical}
non-abelian Yang-Mills field which is averaged over with a Gaussian
distribution. The inelastic $qA$ cross section is then given by~\cite{djm2} 
\be \label{dsigd2b}
q^+ \frac{d\sigma^{qA\to qX}}{dq^+d^2q_t d^2b} &=& \frac{q^+}{P^+} \delta\left(
\frac{p^+ - q^+}{P^+}\right) \frac{1}{(2\pi)^2} C(q_t)~,\non
C(q_t) &=& \int d^2r_t \, e^{i\vec{q}_t\cdot \vec{r}_t}
\exp\left[-2Q_s^2 \int \frac{d^2 l_t}{(2\pi)^2}\frac{1}{l_t^4}
\left(1-\exp(i \vec{l}_t \cdot \vec{r}_t)\right)\right]~.
\ee
(We define $Q_s^2$ as in~\cite{gelis}, not~\cite{djm2}.)
The integrals over $l_t$ are cut off in the infrared by a cutoff
$\Lambda$, which we assume is on the order of $\Lambda_{\rm QCD}$.
At asymptotically large $q_t\gg Q_s$, the correlator $C(q_t)$ smoothly matches
leading-twist perturbation theory: $C(q_t)\to Q_s^2/q_t^4$~\cite{gelis}.
For $q_t\lton Q_s$, the cross section is much flatter than that from
leading-twist perturbation theory~\cite{djm2,gelis}. Thus,
the incident collinear quarks typically are scattered to large transverse
momenta of order $Q_s\gg \Lambda_{\rm QCD}$. This is rather different from the
behavior of leading-twist pQCD, where the momentum transfer peaks
at $q_t\sim\Lambda_{\rm QCD}$.
In eq.~(\ref{dsigd2b}), $Q_s^2$ is in principle a function of $x$, and of
the impact parameter $b$. In what follows, we shall consider only central
collisions where $Q_s^2$ is largest. Also, we focus on a not very large
rapidity interval in the forward region, and so assume that $Q_s$ is
essentially constant.

To compute inclusive production of hadrons, the quark distribution is
convoluted with the (leading twist) quark distribution function of a proton and
a fragmentation function:
\be \label{eq3}
z \frac{d\sigma^{pA\to hX}}{dz d^2k_td^2b}
=\frac{1}{(2\pi)^2}\int d^2q_td^2l_t\int\limits_z^1 dx\frac{x}{z}
f_{q/p}(x,Q_s^2)
D_{q/h}\left(\frac{z}{x},l_t,Q_s^2\right) \delta^2(xl_t/z+q_t-xk_t/z) 
C(q_t)~,
\ee
Note that here we make the natural assumption, in line with the discussion
above, that the
projectile partons are resolved at the scale $Q^2=Q_s^2$, i.e.\ that the
factorization scale is given by the saturation momentum of the dense
target~\cite{Qsfact}. In eq.~(\ref{eq3}),
$k_t$ denotes the transverse momentum of the produced hadron,
and $\log z$ is its rapidity relative to the proton beam; $l_t$ is the
transverse momentum acquired in the fragmentation.
We employ the CTEQ5L parameterization of the parton distribution
functions in the proton~\cite{cteq}.

\begin{figure}[htp]
\centerline{\hbox{\epsfig{figure=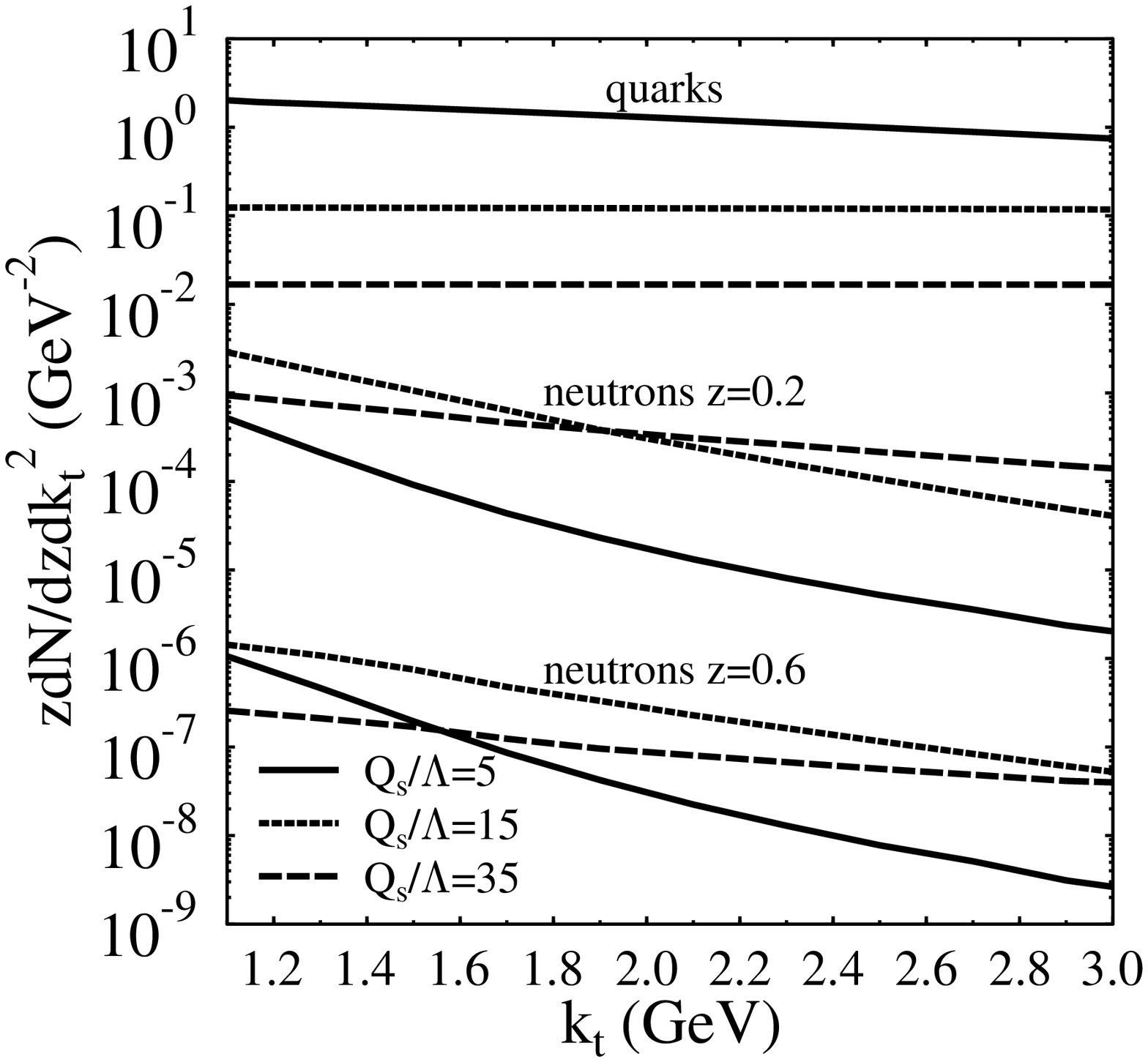,height=8cm}
\epsfig{figure=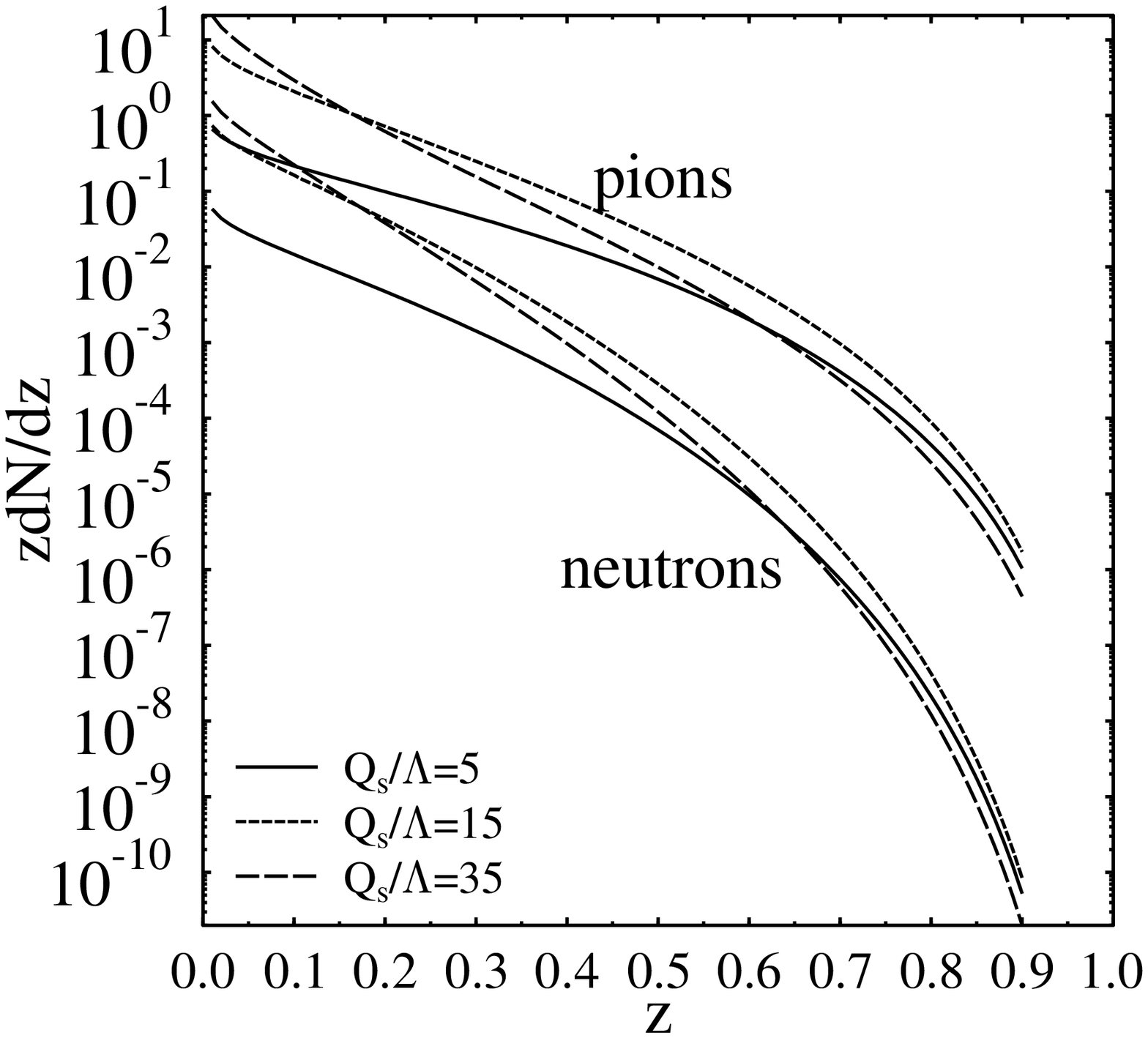,height=8cm}}}
\vspace*{-1cm}
\caption{Left: Transverse momentum
distribution of neutrons (in fact,
$n/2+\bar{n}/2$) from the breakup of an incident proton
at various longitudinal momentum fractions $z$ and target saturation
momenta $Q_s$ (bottom six curves). The top three curves depict the
underlying quark distributions. Right: Longitudinal distributions of
$n/2+\bar{n}/2$ and $\pi^0$, integrated over $k_t$.}
  \label{fig1}
\end{figure}
Since the transverse spread in the fragmentation of the quark 
at $Q_s$ is much smaller than that originating from the scattering on the
dense target, we can safely neglect this effect and use the
fragmentation functions integrated over transverse momentum,
\be
z \frac{d\sigma^{pA\to hX}}{dz d^2k_td^2b}
=\frac{1}{(2\pi)^2}\int\limits_z^1 dx \, \frac{x}{z}f_{q/p}(x,Q_s^2)
D_{h/q}\left(\frac{z}{x},Q_s^2\right)  C\left(\frac{x k_t}{z}\right)~.
\ee
Specifically, we employ the KKP parameterizations~\cite{KKP} for the
fragmentation functions. 
The main contribution to the leading hadron yield comes from $x\sim z$.
Thus, the transverse spectra of the leading hadrons (especially those
of baryons) are tracking very closely the spectra of quarks. 

We give our predictions for the $k_t$ spectra in Fig.~\ref{fig1}. 
As expected, we find that the transverse momentum distribution
at large $z$ flattens as the target density ($\sim Q_s$) increases.
At the same time, the longitudinal ($z$-) distribution steepens,
resulting in larger suppression of forward hadron production.
If we integrate over $k_t$,
we are left with a convolution of the quark distribution in
the proton with the fragmentation function,
$z {d\sigma^{pA\to hX}}/{dz d^2b}
= \int_z^1 dx \, ({z}/{x}) f_{q/p}(x,Q_s^2)
D_{h/q}({z}/{x},Q_s^2){d\sigma^{qA\to qX}}/{d^2b}$,
similar to the expression given in~\cite{Berera:1996ku,foot1},
times the inelastic $qA$ cross section~\cite{djm2}
($=\pi R_A^2$ for a perfectly ``black disc'').
Thus, we obtain an analytical expression for the ``limiting curve''
of hadron production from QCD in the high-density limit. The existence of such
a ``limiting fragmentation curve''~\cite{Benecke:sh}
was confirmed recently by the PHOBOS collaboration at RHIC~\cite{PHOBOS}.
Note, however, that the ``limiting fragmentation curve'' keeps evolving
with $Q_s$, even in the black body limit $Q_s\gg\Lambda$. This is
due to QCD scaling violations, i.e.\ DGLAP evolution of the dilute projectile
fields and of the fragmentation functions.
The ``limiting'' distribution of net baryon number is given by
\be
z \frac{d\sigma^{pA\to (B-\overline{B})X}}{dz d^2b}
=  \int\limits_z^1 dx \, \frac{z}{x}\left[f_{q/p}(x,Q_s^2)
-f_{\bar q/p}(x,Q_s^2)\right]
\left[D_{B/q}\left(\frac{z}{x},Q_s^2\right)
-D_{\overline{B}/q}\left(\frac{z}{x},Q_s^2\right)\right]
\frac{d\sigma^{qA}}{d^2b}~.
\ee
The evolution of the moments of this distribution with $Q_s^2$ in the
black body limit is given by
\be \label{moments}
\frac{d}{d\log Q_s^2} \langle z^n\rangle_{B-\overline{B}} 
= \frac{\alpha_s(Q_s)}{\pi}
\, \gamma^{(0)}_{qq}(n+1)\, \langle z^n\rangle_{B-\overline{B}}~,
\ee
with $\gamma^{(0)}_{qq}$ the LO anomalous dimension for the quark distribution
and fragmentation functions.
\begin{figure}[htp]
\centerline{\hbox{\epsfig{figure=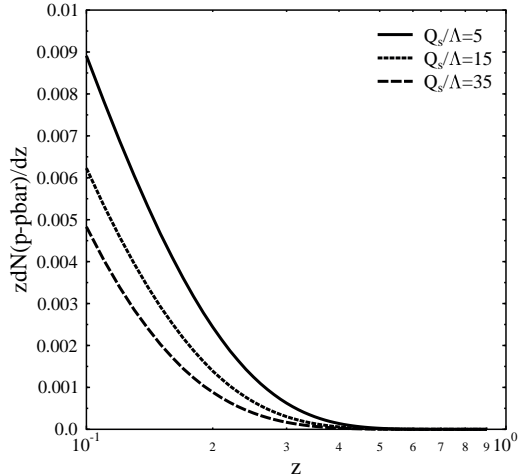,height=8cm}}}
\vspace*{-1cm}
\caption{Longitudinal distribution of net protons.}
  \label{fig2}
\end{figure}
In fig.~\ref{fig2} we show $zdN(p-\overline{p})/dz\equiv
(zd\sigma^{pA}/dzd^2b)/(d\sigma^{qA}/d^2b)$. Lacking a parametrization of
the net proton fragmentation function, we employ the ansatz
$(D_{B/q}(z) - D_{\overline{B}/q}(z)) / D_{B/q}(z)\simeq\sqrt{z}$,
which appears consistent with the data of ref.~\cite{Arneodo:1989ic}
on the $\overline{p}/p$ ratio. This is of course a rough qualitative ansatz
which should not be used at small $z$. In any case, one does not expect that
the assumption of independent fragmentation applies at small $z$,
when the (longitudinal and transverse) momentum degradation in the
fragmentation process is very large. This restricts the applicability of
eq.~(\ref{moments}) to sufficiently high $n$, say $\gsim5$.

Because of the softer fragmentation function of
a quark to a baryon than to a meson we expect that in the ``black-body'' limit 
the leading particle spectrum will be dominated by production of
mesons rather than nucleons, in qualitative difference to $pp$ 
scattering (see Fig.~\ref{fig1}, right).
Hence, though it would be possible to observe the $k_t$ broadening using a 
generic small angle neutron calorimeter, it will be necessary to
do additional measurements to separate the $\pi^0,K_L,\eta,..$ and neutron
contributions. The current upgrade of STAR to extend the acceptance to 
larger rapidities and the BRAHMS detector will provide other opportunities
at the BNL-RHIC accelerator.

Since the suppression of leading hadrons
is so strong one may ask whether other mechanisms
may compete. One such mechanism is the propagation of the nucleon through
the target in a nearly
point-like configuration. If its size is smaller than $1/Q_s$ it will not 
be resolved by the nucleus. However, this mechanism is
power suppressed~\cite{Berera:1996ku}, and strongly peaked at small 
$k_t$. Also, the experiments with the deuteron beam at RHIC can measure
directly the probability for a nucleon to go through the center of the target
without interacting~\cite{FS91}. Hence, this contribution can be
estimated, in principle.

In summary, we argue that for very high energy $pA$ collisions
the production of leading
hadrons, most notably that of forward baryons and anti-baryons, should be
computable in weak coupling QCD. This is because the large gluon density
of the ``black body'' target per unit transverse area at small $x$ provides
an intrinsic semi-hard momentum scale. Resumming higher-twist contributions
to the quark-nucleus cross section we predict a depletion of forward hadron
production, which becomes stronger as the atomic number of the target
nucleus and/or the energy increase. (This should be even more pronounced for
$\Lambda$ hyperons than for nucleons because incident $u$, $d$ quarks
are favored.) At the same time, the transverse momentum distribution of forward
hadrons broadens.

{\bf Acknowledgement:}\\
We thank Leonid Frankfurt for helpful discussions, Francois Gelis for providing
a code to compute the function $C(q_t)$ numerically, and Miklos Gyulassy and
Larry McLerran for drawing our attention to the net proton distribution.
A.D.\ gratefully acknowledges support by the U.S.\ Department of Energy
under contract number No.\ DE-AC02-98CH10886; M.S.\ from contract number
DE-FG02-93-ER-40771. L.G.\ thanks the Minerva Foundation for support.


\begin{thebibliography}{99}
\bibitem{Qsfact}
At present, there is no solid proof for this from full QCD in the
black-body limit. However, for the case of $\gamma^* A$ scattering arguments
in favor of this assumption were given in~\cite{bbl}, as the wave function of
the photon entered at the scale $M^2 \sim Q_s^2$.
\bibitem{foot1}
Except that
in~\cite{Berera:1996ku} the factorization scale was smaller
because the transverse momenta acquired by the partons were 
taken to be substantially smaller than the ones which emerge in the saturation
scenario.
\end{thebibliography}
\end{document}